\documentclass[10pt,a4paper,twocolumn]{article}

\usepackage[T1]{fontenc}
\usepackage[english]{babel}
\usepackage{lmodern}
\usepackage{microtype}
\usepackage{geometry}
\usepackage{amsmath,amssymb,mathtools}
\usepackage{booktabs,longtable,tabularx,array,multirow}
\usepackage[table]{xcolor}
\usepackage{enumitem}
\usepackage{graphicx}
\usepackage{fancyhdr}
\usepackage{titlesec}
\usepackage{natbib}
\usepackage{hyperref}
\usepackage{url}
\makeatletter
\g@addto@macro\UrlBreaks{\do a\do b\do c\do d\do e\do f\do 0\do 1\do 2\do 3\do 4\do 5\do 6\do 7\do 8\do 9}
\makeatother
\usepackage{caption}
\usepackage{ragged2e}

\definecolor{geoblue}{HTML}{153A5B}
\definecolor{geoteal}{HTML}{167D7F}
\definecolor{geolight}{HTML}{EDF4F6}
\definecolor{geogray}{HTML}{5E6872}
\definecolor{geowarn}{HTML}{8A5A00}
\definecolor{geored}{HTML}{8B2F2F}

\hypersetup{
  colorlinks=true,
  linkcolor=geoblue,
  citecolor=geoteal,
  urlcolor=geoblue,
  pdftitle={Optimizing Visibility in Generative Engines: A Critical Survey of Generative Engine Optimization (2023--2026)},
  pdfauthor={Olivier Martinez},
  pdfsubject={Critical survey of Generative Engine Optimization},
  pdfkeywords={GEO, generative engine optimization, AI search, citations, RAG, visibility}
}

\titleformat{\section}{\large\bfseries\color{geoblue}}{\thesection}{0.55em}{}
\titleformat{\subsection}{\normalsize\bfseries\color{geoteal}}{\thesubsection}{0.55em}{}
\titleformat{\subsubsection}{\small\bfseries\color{geoblue}}{\thesubsubsection}{0.5em}{}
\titlespacing*{\section}{0pt}{1.7ex plus .4ex minus .2ex}{0.65ex}
\titlespacing*{\subsection}{0pt}{1.3ex plus .3ex minus .2ex}{0.45ex}

\setlist[itemize]{leftmargin=1.15em,itemsep=1.5pt,topsep=2.5pt}
\setlist[enumerate]{leftmargin=1.35em,itemsep=1.5pt,topsep=2.5pt}
\newcolumntype{Y}{>{\RaggedRight\arraybackslash}X}
\newcolumntype{L}[1]{>{\RaggedRight\arraybackslash}p{#1}}

\newcommand{\PAWC}{\textsc{pawc}}
\newcommand{\peer}{\textcolor{geoteal}{\textbf{PEER}}}
\newcommand{\forth}{\textcolor{geoblue}{\textbf{FORTHC.}}}
\newcommand{\preprint}{\textcolor{geowarn}{\textbf{PREPR.}}}
\newcommand{\workshop}{\textcolor{geogray}{\textbf{WKSP.}}}

\title{Optimizing Visibility in Generative Engines: A Critical Survey of Generative Engine Optimization (2023--2026)}
\author{Olivier Martinez}
\date{July 15, 2026}

\makeatletter
\renewcommand{\@maketitle}{%
  \newpage
  \null
  \vspace*{-1.4em}
  \begin{center}
  {\small\bfseries\color{geoteal}CRITICAL LITERATURE SURVEY \textbullet{} 2023--2026\par}
  \vspace{0.45cm}
  {\LARGE\bfseries\color{geoblue}Optimizing Visibility in Generative Engines\par}
  \vspace{0.18cm}
  {\large A Critical Survey of Generative Engine Optimization\par}
  \vspace{0.5cm}
  {\normalsize \@author\par}
  \vspace{0.08cm}
  {\footnotesize\href{mailto:olivier.martinez2@sciencespo.fr}{olivier.martinez2@sciencespo.fr} \quad
  \href{https://orcid.org/0009-0009-3495-5458}{ORCID 0009-0009-3495-5458}\par}
  \vspace{0.42cm}
  \begin{minipage}{0.94\textwidth}
  \footnotesize
  \colorbox{geolight}{\parbox{0.97\textwidth}{\textbf{Version note.} This survey covers work made public or formally accepted from the first version of the foundational paper, released on November 16, 2023, through July 14, 2026. Peer-reviewed articles, accepted work not yet presented, and preprints are distinguished throughout.}}
  \end{minipage}
  \end{center}
  \vspace{0.55cm}
}
\makeatother

\begin{document}

\maketitle
\thispagestyle{fancy}

\selectlanguage{english}

\begin{abstract}
Generative Engine Optimization (GEO) seeks to increase content's presence, likelihood of citation, or influence in answers produced by generative engines. Since the foundational GEO paper, the field has expanded rapidly, but terminology, metrics, and evidence standards remain heterogeneous. This critical survey reviews 45 studies selected under a November 2023--July 2026 publication window, including one earlier preprint published at EMNLP after the window opened, plus relevant RAG and evaluation work. We argue that GEO is not a single ranking task but a stochastic, partially observable pipeline spanning search activation, crawling and indexing, retrieval, reranking and context allocation, citation, prominence, factual absorption, fidelity, and user behavior. The foundational paper's widely cited gains are valid within its experimental setting but conditional on a source already being present in a fixed context; they establish neither organic discoverability nor durable traffic effects. Reviewed work indicates that topical relevance and context position are the most reproducible levers, generic heuristics transfer poorly, competition can erode individual gains, and citation-oriented rewrites can impair retrieval. Commercial audits further reveal low source overlap, substantial run-to-run variability, and persistent fidelity gaps. We contribute a multistage formal model, a visibility vector separating discoverability, citation, absorption, and economic outcomes, an evidence hierarchy, and a reproducible protocol based on repeated measurements, paraphrases, controls, human validation, and multi-actor interference. Within this corpus, the evidence is narrow: already-retrieved content can causally alter its citation or use, but no reviewed technique shows a stable, longitudinal, cross-platform causal effect on organic discoverability or downstream behavior.
\end{abstract}


\begingroup
\footnotesize
\setlength{\parskip}{0pt}
\renewcommand{\baselinestretch}{0.90}\selectfont
\tableofcontents
\endgroup
\section{Introduction}

Search interfaces built on large language models no longer merely rank a list of links. They retrieve documents, condense them, answer in natural language, and may attach citations to particular claims. This transformation changes the nature of visibility: a publisher no longer seeks only to occupy a position on a results page, but also to be retrieved, used, cited, featured in a prominent passage, and, ideally, selected by the user. The term \emph{Generative Engine Optimization} gave this problem a name and an initial experimental protocol \citep{aggarwal2024geo}.

The scientific significance is twofold. First, generative engines are becoming intermediaries for access to information. Their source-selection decisions distribute attention, authority, and revenue. Second, conversational interfaces make the allocation of visibility less transparent than conventional ranking: the same source may be absent, cited without being used, paraphrased without a link, or decisive in shaping the structure of an answer. Mention frequency alone is therefore insufficient.

Commercial claims about GEO, however, have advanced faster than the evidence. The \emph{up to 40\%} figure from the foundational paper is often recast as a general promise of ranking highly in ChatGPT, although it describes a relative visibility gain in a simulator in which five documents have already been placed in context. Conversely, the contrary findings of \citet{puerto2025cseo} are sometimes presented as a refutation of GEO, even though they evaluate a different outcome across several domains and multiple actors. The two bodies of evidence can be reconciled once the stages of the pipeline and the causal estimand are distinguished.

This survey advances four propositions.

\begin{enumerate}
  \item \textbf{GEO is multistage.} Discoverability, rank within the context, citation, prominence, absorption, and traffic are distinct variables. An intervention may improve one stage while impairing another.
  \item \textbf{Visibility is a distribution.} It depends on the engine, date, location, query formulation, whether search is actually activated, and stochasticity in generation. A point estimate is not a stable indicator.
  \item \textbf{The evidence is primarily conditional on retrieval.} Experiments provide strong evidence that a document that has already been retrieved can alter an answer. They establish far less often that a page will be retrieved organically, and almost never that it will produce a durable effect on conversions.
  \item \textbf{Optimization and manipulation share a channel.} Helpful rewrites, adversarial sequences, and indirect prompt injections all modify the text consumed by the engine. The normative distinction should rest on truthfulness, semantic preservation, disclosure, and the absence of hidden instructions, not solely on the objective of visibility.
\end{enumerate}

Our contributions are a structured review of 45 studies; a formalization of the pipeline; a taxonomy of metrics and interventions; a critical assessment of evidence from commercial engines; an account linking competitive optimization, security, and the creator economy; and, finally, a minimum protocol for measurement and a research agenda.

\section{Scope and Review Method}

\subsection{Review Type and Cutoff Date}

This article is a \emph{critical scoping review} rather than a systematic review in the clinical sense. The field has not yet converged on a stable vocabulary, many studies remain available only as preprints, and the systems under study change over the publication cycle. Our objective is therefore to map the research landscape, assess the strength of the inferences, and propose a cumulative framework, without claiming to meta-analyze incomparable effect sizes.

The primary review window begins on November 16, 2023, the date of the first arXiv version of \citet{aggarwal2024geo}, and ends on July 14, 2026. We include \citet{liu2023verifiability}, whose preprint predates this starting point but whose publication in the EMNLP proceedings, in December 2023, appeared after that date. This study provides an important contemporaneous benchmark for citation fidelity.

\subsection{Search, Selection, and Coding}

The search covered arXiv, the ACM Digital Library, ACL Anthology, the NeurIPS proceedings, PMLR, and OpenReview, followed by backward and forward citation searching from the core studies. Search-term families included \emph{generative engine optimization}, \emph{answer engine optimization}, \emph{conversational SEO}, \emph{AI search visibility}, \emph{citation absorption}, \emph{ranking manipulation}, \emph{AI Overview citations}, and the names of commercial engines.

A study was included if it met at least one of the following criteria:

\begin{itemize}
  \item it defines or evaluates an intervention targeting the retrieval, rank, citation, mention, or influence of a source;
  \item it measures source visibility, stability, fidelity, or concentration in a commercial generative engine;
  \item it studies an attack or defense directly involving the ranking or use of documents by a conversational engine;
  \item it formalizes the incentives and externalities created by the distribution of generative visibility.
\end{itemize}

Generic work on RAG was excluded unless it contributed a directly necessary metric, such as subquestion coverage \citep{xie2025coverage}. Each study was coded by publication status, pipeline stage, whether the system was controlled or commercial, unit of measurement, principal finding, and most consequential threat to validity. The complete matrix and a retrospective record of the search protocol are provided as arXiv ancillary files; a condensed version appears in the appendix.

\subsection{Publication Status and Interpretive Caution}

The corpus combines peer-reviewed articles, formally accepted but forthcoming papers, workshop papers, and preprints. This distinction is substantive, not cosmetic. As of July 14, 2026, the studies by \citet{grossman2026disrupts} and \citet{vishwakarma2026cited} have been accepted to SIGIR 2026, but the conference begins on July 20; they are therefore described as \emph{forthcoming}. Several prominent measurement studies have not been peer-reviewed \citep{schulte2026measure,zhang2026absorption,xu2026aio,allaham2026synthetic}. We use their data where they complement published findings, but do not assign them the same evidentiary weight.

\subsection{Review Limitations}

The pace of the field creates a risk of misalignment between a manuscript, the product it audits, and the interface available to the reader. Commercial names also conceal different configurations: an API receiving URLs, a model equipped with a web tool, and a consumer-facing interface are not the same system. The original search did not retain database-specific hit counts or a complete exclusion ledger; the ancillary protocol records this limitation rather than reconstructing unavailable data. Finally, task heterogeneity precludes a meaningful aggregation of percentage gains. Our synthesis therefore emphasizes the direction of effects, their conditions, and their level of evidence rather than an artificial average, and any claim of absence is bounded to the reviewed corpus.

\section{From Generative Engines to GEO: A Multistage Formalization}

\subsection{A Stochastic and Partially Observable Pipeline}

Let $q$ denote a query, $u$ a user state (language, location, history, account), $e$ an engine, $t$ a point in time, $W_t$ the accessible web, and $\varepsilon$ a randomness term. An engine may first decide whether to activate search:
\begin{equation}
  A = g_{\mathrm{act}}(q,u,e,t,\varepsilon_A) \in \{0,1\}.
\end{equation}
If search is activated, the engine retrieves a set $R$ and then ranks or reranks it:
\begin{align}
  R &= g_{\mathrm{ret}}(q,u,W_t,e,\varepsilon_R),\\
  \pi(R) &= g_{\mathrm{rank}}(q,R,u,e,t,\varepsilon_K).
\end{align}
The generator produces an answer $Y$ and a set of citations $C$ from a context window $\pi(R)_{1:k}$:
\begin{equation}
  (Y,C) = g_{\mathrm{gen}}(q,\pi(R)_{1:k},u,e,t,\varepsilon_Y).
\end{equation}
The user may then read the answer, click a link, complete a conversion, or ignore it. Content creators do not generally observe the complete set $R$, the reranking score, or the generator's internal states. GEO is therefore a black-box optimization problem under incomplete information.

\begin{figure*}[t]
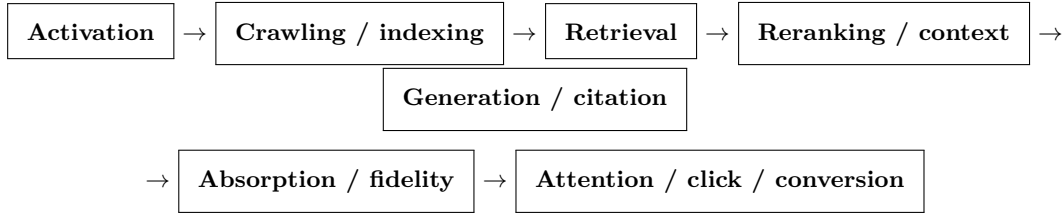

\centering
\setlength{\fboxsep}{7pt}
\small
\fbox{\textbf{Activation}} $\rightarrow$
\fbox{\textbf{Crawling / indexing}} $\rightarrow$
\fbox{\textbf{Retrieval}} $\rightarrow$
\fbox{\textbf{Reranking / context}} $\rightarrow$
\fbox{\textbf{Generation / citation}}\\[6pt]
$\rightarrow$ \fbox{\textbf{Absorption / fidelity}} $\rightarrow$
\fbox{\textbf{Attention / click / conversion}}
\caption{The causal visibility pipeline. Most GEO studies optimize the stages between context allocation and citation; far fewer observe crawling, organic retrieval, or user behavior.}
\label{fig:pipeline}
\end{figure*}

\subsection{A Visibility Vector Rather Than a Single Rank}

For a source $s$, we propose the vector
\begin{equation}
\mathbf{V}_s = (D_s,K_s,C_s,P_s,H_s,F_s,B_s),
\end{equation}
where:
\begin{itemize}
  \item $D_s$ is retrieval probability, or discoverability;
  \item $K_s$ describes exposure in the context (rank, top-$k$ inclusion, allocated tokens);
  \item $C_s$ is the probability of a mention or citation;
  \item $P_s$ is observable prominence (position, repetition, attributed share);
  \item $H_s$ is absorption, namely the source's effective contribution to the facts, language, or structure of the answer;
  \item $F_s$ is fidelity, the extent to which attributed claims are actually supported and rendered accurately;
  \item $B_s$ is the behavioral or economic outcome (click, referral, conversion, value).
\end{itemize}

A scalar score $M_w=\mathbf{w}^{\top}\mathbf{V}_s$ is defensible only when the weights $\mathbf{w}$ correspond to an explicit objective. Aggregating a mention, an accurate citation, and a conversion without a utility model merely obscures normative choices. Moreover, $D_s$ and $C_s$ should be reported separately: a high conditional probability of citation, $\Pr(C_s=1\mid s\in R)$, does not compensate for a low probability of retrieval.

\subsection{The Causal Estimand of an Intervention}

Let $T$ be a transformation of a document $d_s$, and let $m$ be a metric. The average treatment effect of interest is
\begin{equation}
\tau_T(m)=\mathbb{E}\!\left[m\{Y(T(d_s))\}-m\{Y(d_s)\}\right],
\label{eq:ate}
\end{equation}
with the query, engine, corpus, context order, and timing controlled or randomized. This notation highlights three challenges. First, an unpaired before--after comparison confounds the treatment with stochasticity in generation. Second, holding $R$ fixed estimates an effect \emph{conditional} on retrieval, not a total effect on the pipeline. Third, the no-interference assumption is violated: when one source gains normalized share, another loses it, and when all competitors optimize, one actor's treatment changes the outcomes of the others.

\section{The Foundational Paper: Contributions, Results, and Empirical Scope}

\subsection{What the Paper Actually Established}

\citet{aggarwal2024geo} introduced three elements that continue to structure the field. They named the phenomenon, modeled a generative engine as a combination of retrieval and synthesis, and proposed metrics suited to citations dispersed throughout a text. Their GEO-bench contains 10\,000 queries drawn from several datasets. For each query, the top five Google results are provided to GPT-3.5-turbo; five answers are generated at temperature 0.7. One source is modified under each of nine alternative strategies and then compared with its original version.

The metrics are (i) \emph{Word Count}, the share of words attributed to a source; (ii) \emph{Position-Adjusted Word Count} (\PAWC), which discounts passages appearing later in the answer; and (iii) \emph{Subjective Impression}, an LLM judgment inspired by G-Eval \citep{liu2023geval}. The shares assigned to the five sources sum to one. Visibility is therefore intrinsically relative and redistributive.

\begin{table*}[t]
\centering
\small
\caption{Principal results from \citet{aggarwal2024geo}. Values are mean percentage shares, normalized to sum to 100 across sources.}
\label{tab:founder}
\begin{tabular}{lrrl}
\toprule
Intervention & \PAWC & Subjective Impression & Critical interpretation \\
\midrule
Baseline & 19.3 & 19.3 & Reference condition \\
Keyword stuffing & 17.7 & 20.2 & Reduces the position-adjusted metric \\
Fluency & 24.7 & 21.9 & Moderate, domain-dependent gain \\
Cite Sources & 24.6 & 21.9 & Helps within the fixed context \\
Quotation Addition & 27.2 & 24.7 & Highest \PAWC; approximately 41\% relative gain \\
Statistics Addition & 25.2 & 23.7 & Substantial gain, factuality not guaranteed \\
\bottomrule
\end{tabular}
\end{table*}

The \emph{up to 40\%} figure derives primarily from the increase in \PAWC from 19.3 to 27.2 for Quotation Addition, or approximately 41\% in relative terms. The largest gain in Subjective Impression is lower. This result does not mean that 40\% more readers will click, nor that a page will gain 40\% in retrieval probability. It means that, in this testbed, a source already provided to the generator receives a larger position-weighted share of attributed text.

\subsection{Substantive Findings}

Three observations have held up well in the subsequent literature. First, keyword stuffing does not transfer effectively from conventional SEO. Second, directly extractable information---figures, definitions, quotations, and references---can facilitate the use of a document. Third, effects depend on the domain and initial rank. Under the \emph{Cite Sources} strategy, the fifth source gains 115.1\% while the first loses 30.3\%, illustrating the competitive nature of the metric.

The Perplexity test provides useful but limited validation: it includes 200 examples, with texts uploaded as files rather than web pages retrieved organically. The largest reported gains reach 22\% on \PAWC and 37\% on Subjective Impression. It is therefore a black-box test of document use, not an experiment in crawling or ranking in production.

\subsection{Structural Limitations}

The principal limitations are not errors in the paper; rather, they define the agenda it left open:
\begin{itemize}
  \item the source is already present in a five-document context, so $D_s$ and much of $K_s$ are fixed;
  \item without a user study, \PAWC assumes that an earlier citation receives more attention;
  \item the subjective judge is an LLM from the same model family, and its poorly calibrated score is renormalized to the \PAWC distribution;
  \item interventions that add statistics or references are not subject to a strong truthfulness constraint;
  \item no clicks, referrals, traffic, or purchases are observed;
  \item the engine and benchmark are snapshots, whereas deployed products change rapidly.
\end{itemize}

The paper's enduring contribution therefore lies less in providing a recipe than in transforming a commercial intuition into an experimental problem. The subsequent literature can be read as a progressive decomposition of the variables that the paper had grouped under \emph{impression}.

\section{Evolution of the Field, 2023--2026}

\begin{table*}[t]
\centering
\small
\caption{Four successive shifts in GEO research.}
\label{tab:waves}
\begin{tabularx}{\textwidth}{L{1.45cm}L{2.8cm}YY}
\toprule
Period & Central question & Representative studies & Conceptual shift \\
\midrule
2023--2024 & Can a source gain visibility? & \citet{aggarwal2024geo}; \citet{wan2024evidence} & From ranked links to answer share; first controlled interventions \\
2024--2025 & Can rankings and recommendations be manipulated? & \citet{pfrommer2024ranking}; \citet{nestaas2025adversarial}; \citet{kumar2024manipulating} & Retrieved content becomes an attack surface; white-hat/adversarial distinction \\
2025 & Do heuristics generalize under competition? & \citet{puerto2025cseo}; \citet{qian2025rankingblind}; \citet{chen2025dominate} & Countervailing evidence, multi-actor congestion, and cross-engine differences \\
2026 & How can we measure the full pipeline and learn optimization policies? & \citet{kim2026sageo}; \citet{liu2026featgeo}; \citet{kirsten2026characterizing}; \citet{schulte2026measure} & Stage-specific optimization, absorption metrics, longitudinal audits, and governance \\
\bottomrule
\end{tabularx}
\end{table*}

This chronology is not strictly sequential, but it reveals a maturing field. The first shift isolates content transformations. The second shows that the same channel can subvert a recommendation. The third adds questions of generalizability, domain variation, and competition. The fourth moves upstream toward retrieval, downstream toward traffic, and treats visibility as an unstable distribution.

\section{Measuring Visibility: From Mentions to Effects}

\subsection{A Hierarchy of Metrics}

The literature uses the term ``visibility'' to refer to at least nine distinct quantities. Table~\ref{tab:metrics} orders them from the easiest signal to observe to the outcome most closely tied to user value. Each level addresses a distinct question and has its own denominator.

\begin{table*}[t]
\centering
\small
\caption{Taxonomy of visibility metrics.}
\label{tab:metrics}
\begin{tabularx}{\textwidth}{L{2.4cm}L{3.15cm}YY}
\toprule
Level & Example metric & What it identifies & What it does not establish \\
\midrule
Activation & $\Pr(A=1\mid q)$ & Probability that the generative surface or search is activated & Visibility of a particular source \\
Retrieval & recall@$k$, URL/domain presence & Discoverability in the candidate pool or context & Citation or influence in the response \\
Mention & presence of a brand or entity & Minimal nominal exposure & Tone, source, evidence, or click \\
Citation & citation rate; rank of first citation & Explicit selection as a source & Exact support or substantive contribution \\
Prominence & word count; \PAWC; repetition & Observable share and position & Human attention without behavioral validation \\
Coverage & share of subquestions or units covered & Breadth of supported information needs & Causal effect of the source on wording \\
Absorption & similarity, ablation with/without the source, composite score & Contribution to facts, language, or structure & Internal model states when the score remains a proxy \\
Fidelity & citation recall/precision; entailment & Alignment between a claim and its source & Positive visibility or an economic outcome \\
Behavior & click, referral, conversion, revenue & Observable downstream outcome & Causality without a control group or baseline trend \\
\bottomrule
\end{tabularx}
\end{table*}

The \emph{citation recall} metric of \citet{liu2023verifiability} measures the share of verifiable sentences that are fully supported; \emph{citation precision} measures the share of citations that correctly support their associated sentence. Across Bing Chat, NeevaAI, Perplexity, and YouChat, only 51.5\% of sentences were fully supported, and 74.5\% of citations supported the proposition with which they were associated. These metrics evaluate engine fidelity rather than publisher visibility, but they impose an essential constraint: a visible citation may be incorrect.

Coverage metrics address a different limitation. A response may cite many domains while omitting important subquestions. \citet{xie2025coverage} decompose complex information needs into subquestions; \citet{huang2026bubbles} distinguish the number of sources from the fraction of cited-source content reflected in the summary's atomic content units. This family of metrics brings visibility closer to informational utility.

Finally, the notion of ``absorption'' proposed by \citet{zhang2026absorption} distinguishes selection from contribution: ChatGPT may cite fewer sources while relying more heavily on each, whereas Google or Perplexity distribute their citations more broadly. The concept is productive, but the published score combines position, repetition, coverage, and textual similarity. It does not reveal an internal causal trace. A genuine measure of absorption would ideally require a matched ablation with and without the source, followed by an analysis of the facts and formulations that change.

\subsection{Reliability: Repetition, Paraphrase, and Time}

A generative engine is repeatable as an experiment only at the distributional level. Even a reported temperature of zero fixes neither the index, nor retrieval, nor the versions of external services. Across four engines and 45 days, \citet{schulte2026measure} observe daily source-level Jaccard scores of approximately 0.34--0.42, with similar levels for repetitions within 24 hours. They propose seven to eight repetitions per prompt as a starting point. This number is not a universal standard: it derives from a small universe of Swiss queries and at most ten repetitions. The appropriate practice is sequential precision analysis: repeat the measurement until the interval around the estimand is sufficiently narrow for the decision at hand.

Peer-reviewed audits confirm this phenomenon. \citet{kirsten2026characterizing} analyze 4\,706 queries across several surfaces in the United States and Germany. Page overlap across two months is 18\% for AI Overviews, compared with 45\% for organic Google; on the surfaces for which temperature could be controlled, repeated runs at temperature zero change 9--28\% of decisions. \citet{grossman2026disrupts} likewise find that minor reformulations alter AIO sources more than those returned by conventional search. A GEO measurement must therefore vary along at least four dimensions: run, paraphrase, date, and engine.

\subsection{Denominators and Missing Outputs}

Discarding responses without search or citations creates selection bias. In the configuration studied by \citet{schulte2026measure}, 57.8\% of ChatGPT repetitions did not activate web search. A dashboard cannot calculate a ``share of citations'' only among responses that contain citations and then interpret it as overall visibility. It must decompose:
\begin{equation}
\begin{aligned}
\Pr(s\ \text{cited})
  &= \Pr(A=1) \\
  &\quad\times \Pr(s\in R\mid A=1) \\
  &\quad\times \Pr(s\ \text{cited}\mid s\in R,A=1).
\end{aligned}
\end{equation}
This simple identity explains why high conditional citation rates can coexist with low commercial visibility.

URL canonicalization is also critical. Parameters, redirects, anchors, AMP versions, translated pages, and aggregators can artificially fragment a domain. Studies should publish their resolution rules and retain the raw URL, final URL, registrable domain, and, where possible, a hash of the retrieved content.

\subsection{LLM Judges and Human Validation}

LLM judges make it possible to evaluate tens of thousands of claims, but they introduce model dependence, stylistic bias, and circularity. The risk is greatest when the same model generates the rewrite, the response, and the score. A robust evaluation should:
\begin{enumerate}
  \item separate the generator and judge model families;
  \item blind the judge to the experimental condition;
  \item randomize the order of variants;
  \item validate each dimension on a stratified human sample;
  \item report agreement, sensitivity, specificity, and disagreement patterns, rather than only an overall correlation;
  \item retain an observable metric that does not rely on a judge, such as a canonical citation or click.
\end{enumerate}

\citet{xu2026aio} report 95.6\% accuracy for their verifier on a manually assessed sample, which strengthens their findings, although inaccessible video or social-media pages remain difficult to evaluate. \citet{vykopal2026credibility} likewise use an automated evaluator for part of their groundedness analysis. These studies point in the right direction: a judge is acceptable as a measured instrument, not as an unexamined source of truth.

\section{Influence Techniques: What Holds Up and What Depends on the Setting}

\subsection{Topical Relevance and Position: The Two Most Robust Factors}

Query--document relevance is the most reproducible factor. In ConflictingQA, \citet{wan2024evidence} use counterfactual interventions to show that models strongly favor explicit alignment with the question, often more than human credibility cues such as scientific references or a neutral tone. This result does not imply that credibility is irrelevant; it indicates that, under conflicting evidence and limited context, the model first follows what appears to answer the question directly.

Rank within the context is equally consequential. \citet{puerto2025cseo} find that moving a source higher in the context has a greater effect than most rewrites. The factorial experiment of \citet{vishwakarma2026cited}, comprising 252\,000 trials across six LLMs and eighteen factors, identifies relevance and position as the primary determinants of the first citation. This finding redirects GEO toward upstream stages: a perfectly formulated page that is absent from the top-$k$ cannot contribute.

\subsection{Extractable Evidence, Structure, and Recency}

Statistics, definitions, comparisons, prices, dates, and references have a plausible advantage: they form units that the generator can select and attribute. The foundational paper, AutoGEO, and FeatGEO observe positive effects for several of these properties \citep{aggarwal2024geo,wu2026autogeo,liu2026featgeo}. The controlled experiment of \citet{vishwakarma2026cited} finds effects for explicit prices and recent dates, whereas formatting changes alone have weak effects.

Two qualifications are necessary. First, the effect depends on intent. A recent date helps with a time-sensitive query, but not necessarily with a stable definition. A quotation may help answer a historical question, yet burden a product listing. Second, adding a fabricated statistic may increase reuse while degrading epistemic quality. The criterion is therefore not to ``add numbers,'' but to provide relevant, verifiable, dated, and properly attributed evidence.

HTML and document structure are beginning to be studied separately. \citet{yu2026structural} report gains associated with certain structural features, but the work's preprint status and reliance on automated judges warrant caution. SAGEO Arena provides a more instructive result: structural fields may improve retrieval without necessarily producing the same effect at the reranking or citation stages \citep{kim2026sageo}. Structure should be evaluated stage by stage, not treated as a universal talisman.

\subsection{General Heuristics Generalize Poorly}

C-SEO Bench provides the principal empirical corrective. Across two tasks, six domains, approximately 1\,900 queries, and 16\,360 documents, only three of 54 method--domain combinations are significantly positive in the main experiment; none is positive in question answering \citep{puerto2025cseo}. Several transformations even reduce rank. Gains decline as adoption increases, ultimately producing congested dynamics that approach a zero-sum game.

E-GEO reaches a compatible conclusion in e-commerce: ten of the fifteen initial heuristics are neutral or negative, whereas systematically optimized prompts perform better; the authors report a relatively stable, domain-agnostic structure \citep{bagga2025egeo}. Recent optimization systems---AutoGEO, FeatGEO, Mind Reader, IF-GEO, and MAGEO---replace fixed recipes with learned rules, intent decomposition, multiple objectives, or a memory of strategies \citep{wu2026autogeo,liu2026featgeo,chen2026mindreader,zhou2026ifgeo,wu2026mageo}. They sometimes achieve substantial gains, but the candidate document sets are generally fixed in advance and inference costs remain high.

\subsection{The End-to-End Test: SAGEO Arena}

SAGEO Arena is crucial because it reinstates retrieval and reranking. Across 171\,003 documents and 2\,700 queries, body-only optimization reduces average top-20 presence by approximately 9\%, top-10 presence after reranking by 16\%, and final citation by 6\% \citep{kim2026sageo}. Applying AutoGEO to the body alone can produce larger losses. A rewrite may therefore perform well once injected while making the document less retrievable or less competitive upstream.

This result resolves an apparent contradiction. Fixed-context benchmarks estimate a direct effect on generation; SAGEO estimates the composition of several effects. If $T$ increases $\Pr(C_s=1\mid s\in R)$ but decreases $\Pr(s\in R)$, the total effect may be negative. Any operational claim must specify which of these effects it measures.

\subsection{What Can Reasonably Be Recommended}

\begin{table*}[t]
\centering
\small
\caption{Level of empirical support for the main white-hat levers.}
\label{tab:levers}
\begin{tabularx}{\textwidth}{L{3.0cm}L{2.6cm}YY}
\toprule
Lever & Level of support & Conditions & Cautious interpretation \\
\midrule
Query--document relevance & Strong in controlled settings & Well-defined intent; no fabrication & Explicitly address genuine information needs \\
Position in context & Strong & Document already retrieved & Highlights the importance of retrieval/reranking \\
Extractable evidence & Moderate to strong & Compatible truthfulness, attribution, and intent & Verifiable figures, definitions, and comparisons \\
Recency, prices, dates & Moderate & Time-sensitive or commercial queries & Useful but non-universal signals \\
Document structure & Moderate and heterogeneous & Distinct effects at each stage & Test headings, tables, and fields without assuming the direction of effect \\
Fluency / simplification & Weak to moderate & Domain- and engine-specific & Optimize for the user first \\
Authoritative tone & Weak and unstable & May conflict with credibility & Do not conflate confidence with evidence \\
Keyword stuffing & Null or negative & Multiple benchmarks & Avoid \\
Formatting alone / fixed recipes & Poor generalization & Occasionally local gains & Requires matched, multi-engine testing \\
\bottomrule
\end{tabularx}
\end{table*}

The best-supported recommendation is conservative: produce a relevant, comprehensive, verifiable, clearly structured, and technically retrievable page; then measure retrieval, citation, and fidelity separately. This strategy more closely resembles high-quality information engineering than keyword manipulation.

\section{Commercial Engines: External Visibility, Instability, and Attribution}

\subsection{Surfaces Do Not Share the Same Sources}

Audits consistently find low overlap. In an audit totaling 1\,008 responses across three systems, the phase-specific analysis of 672 Bing Chat and Perplexity responses identified 355 unique domains, 26\% of which were cited by both systems \citep{li2024arbiters}. \citet{kirsten2026characterizing} observe that 53\% of domains cited by Google AIO do not appear in the organic top 10 and that 27\% are absent from the top 100. Across 11\,500 queries, \citet{grossman2026disrupts} report URL-level Jaccard similarities of 0.11--0.18 among organic Google, AIO, and Gemini.

These results refute the notion of a global GEO ranking. Visibility is indexed by engine and surface. A site visible in the conventional SERP may be absent from AIO; a domain cited by Perplexity may never appear in ChatGPT. Any strategy and metric must therefore identify the product, search mode, and period under study.

\subsection{Activation and Query Profile}

The probability of displaying a generative response depends strongly on query form. \citet{xu2026aio} analyze 55\,393 trending queries over 40 days: the overall AIO activation rate is 13.7\%, but rises to 64.7\% for queries phrased as questions. \citet{grossman2026disrupts} observe AIO for 51.5\% of queries in their representative sample. These figures are not contradictory: the distributions of queries, dates, and categories differ. They illustrate precisely why a rate reported without a description of the sample has limited transportability.

\subsection{Citation Implies Neither Credibility nor Support}

The foundational audit of citation fidelity had already demonstrated the gap between appearance and support \citep{liu2023verifiability}. In 2026, \citet{vykopal2026credibility} observe credible-source shares of 71.4--86.3\%, depending on the assistant and topic, with more misinformation sources in some GPT configurations than in Perplexity or Qwen. \citet{xu2026aio} classify approximately 11\% of 98\,020 atomic claims as insufficiently supported, subject to limitations in retrieval and adjudication. \citet{allaham2026synthetic} report that approximately 16\% of the 19\,154 textual pages successfully retrieved and classified were labeled AI-generated by the selected detector; 27.1\% of URLs were not scraped because they were inaccessible, removed, or non-textual.

Visibility should therefore never be maximized independently of $F_s$. A source may be cited for a claim it does not support, used negatively, or embedded in a synthetic-content feedback loop. Audits should pair citation rates with a matrix covering tone, attribution accuracy, and factual support.

\subsection{From Recognition to Discovery}

\citet{sharma2026discovery} distinguish name recognition from category-level discovery for 112 startups. In their protocol, ChatGPT recognizes 99.4\% of products when they are named, but surfaces them in only 3.32\% of organic discovery queries; Perplexity falls from 94.3\% to 8.29\%. These figures come from a single-study preprint using two models, but the distinction is fundamental. Encoding an entity in model weights or retrieving it by name does not imply recommending it for a generic intent.

The observations of \citet{chen2025dominate} concerning the overrepresentation of \emph{earned media} should be read in this context: third-party mentions may expand the ecosystem of retrievable evidence. The study remains observational and industry-supported; it does not show that securing external coverage mechanically causes a recommendation. It nevertheless suggests that the relevant unit of GEO may be a network of sources rather than an isolated page.

\subsection{Traffic and Conversions: The Weakest Evidence}

Very few studies observe $B_s$. \citet{watanabe2026traffic} analyze logs from a website on which some pages received an AEO intervention. Total ChatGPT referrals increased by a factor of 5.7, but untreated pages had already increased by a factor of 3.5 as the platform grew. A controlled time-series analysis estimates an additional multiplier of 1.82, with a 95\% interval of [1.31, 2.54], while a conservative temporal placebo yields $p=0.16$. The effect is therefore suggestive rather than causally established.

The industry study by \citet{zhang2026pinterest} reports a 20\% production traffic lift versus control following the large-scale deployment of a VLM-agent framework. This is rare production evidence, but group sizes, the unit of assignment, statistical uncertainty, and the components of the intervention are not described in sufficient detail to support a general estimate. At this stage, claims about GEO return on investment clearly outstrip the academic evidence.

\section{Competition, Manipulation, and Defenses}

\subsection{A Technical Continuum and a Normative Boundary}

White-hat optimization and adversarial attacks share the mathematical objective in Equation~\eqref{eq:ate}. What distinguishes them is not whether they exert influence, but the constraints imposed on $T$. We propose four cumulative tests:
\begin{enumerate}
  \item \textbf{semantic preservation}: do the facts and qualifications remain true?
  \item \textbf{evidentiary authenticity}: are statistics, reviews, and references verifiable?
  \item \textbf{content--instruction separation}: does the document inform the user rather than issue hidden commands to the model?
  \item \textbf{disclosure and fairness}: is the commercial intent disclosed, and are competitors represented without fabricated disparagement?
\end{enumerate}

Reorganizing paragraphs or adding a verified primary source will generally satisfy these tests. A model-directed sequence, fabricated testimonial, or instruction to favor a brand will violate them. This framework prevents a rewrite from being classified as ``white-hat'' merely because it is fluent.

\subsection{Evidence of Manipulation in Commercial Systems}

\citet{pfrommer2024ranking} show that indirect injection into a document can raise a target by approximately three ranks on the Perplexity Sonar Large Online API. Because the URLs are supplied explicitly, the evidence concerns the post-retrieval stage. \citet{nestaas2025adversarial} provide the strongest published demonstration across multiple systems: for example, their preference-manipulation attacks increase the recommendation rate of a fictitious camera from 34.0 to 59.4\%, while some plugin selections increase by as much as a factor of 7.2. The products have since evolved and the pages are controlled, but the existence of the vulnerability is established.

Work on rankers reinforces this diagnosis. \citet{qian2025rankingblind} identify a decision-making blind spot, while \citet{xing2026raf} optimize short suffixes that promote items while remaining relatively natural. StealthRank jointly pursues rank and stealth, but its primary status is that of a preprint with a workshop version, not a regular ICML publication \citep{tang2025stealthrank}. CORE reports high Top-1 success rates across four model families, but passes a fixed retrieved list of ten products to the ranker in JSON format, always places the target last, and can generate fabricated reviews \citep{jin2026core}. It measures reranking manipulation, not organic discovery.

\subsection{Competition and Interference}

When impression shares sum to one, GEO is at least locally redistributive. C-SEO Bench shows that gains can erode as adoption increases \citep{puerto2025cseo}. Complementary multi-attacker experiments by \citet{nestaas2025adversarial} demonstrate post-retrieval interference, but do not establish an organic web equilibrium. \citet{hu2025dynamics} models attack choices as an infinitely repeated prisoner's dilemma and studies conditions for cooperation. In a distinct framework, \citet{wu2026ecosystem} model competition among creators and show, under their assumptions, that citation and compensation can sustain content-production effort.

This interference invalidates single-actor evaluations as predictions of equilibrium outcomes. A study should include several saturation levels---for example, 0, 25, 50, 75, and 100\% of documents treated---and estimate both the direct effect and spillover effects. Otherwise, an initial individual gain may fall to zero once the strategy becomes widespread.

\subsection{Defenses}

Defenses cannot be limited to filtering a few suspicious words. In GEO-BENCH, adversarial attacks remain detectable by at least one of the two proxies in most configurations; however, some black-box white-hat rewrites evade both the lexical filter and the perplexity proxy in certain domains \citep{nimase2026geobench}. GRADA uses inter-document relationships for reranking and reduces the success of some attacks by up to about 80\%, with only a limited loss in accuracy \citep{zheng2025grada}. Engines should also separate instructions from retrieved content; limit the authority granted to documents; compare multiple independent sources; detect conflicts of interest and unverifiable evidence; retain attribution logs; and audit effects by actor category.

Defense nevertheless creates a distributional problem. An overly strict anti-promotion filter may penalize small publishers that legitimately describe their products, whereas a permissive filter favors actors able to produce large volumes of content. Effectiveness must therefore be assessed alongside false positives, source diversity, and effects on competition.

\section{Critical Synthesis: What the Literature Actually Establishes}

\begin{table*}[t]
\centering
\small
\caption{Confidence in the field's principal claims.}
\label{tab:claims}
\begin{tabularx}{\textwidth}{L{2.15cm}YL{3.55cm}}
\toprule
Confidence & Claim & Empirical basis and caveat \\
\midrule
High & A document already placed in the context can causally alter its rank, citation, or use. & Controlled replications across multiple models; does not address organic retrieval. \\
High & Query--document relevance and context position are major determinants. & Counterfactual experiments, benchmarks, and a factorial study; effects may vary with length and task. \\
High & Commercial engines differ from one another and vary over time. & Published audits and large preprints; heterogeneous products and periods. \\
High & Retrieved documents constitute a genuine attack surface. & EMNLP 2024, ICLR 2025, and ranker studies; many effects remain post-retrieval. \\
Moderate & Extractable evidence and suitable structure often facilitate use. & Consistent findings, but dependent on intent, engine, and factuality. \\
Moderate & Systematically optimized or learned methods often outperform fixed heuristics in controlled benchmarks. & AutoGEO, E-GEO, FeatGEO, and agentic methods; fixed contexts and substantial costs. \\
Moderate & Competitive adoption can erode individual gains in tested multi-actor settings. & C-SEO Bench; complementary post-retrieval multi-attacker experiments; few real-web ecosystem studies. \\
Low & A white-hat GEO intervention durably improves organic discoverability across multiple engines. & Very few end-to-end tests; SAGEO even finds adverse upstream effects. \\
Very low & Citation scores predict clicks, conversions, or revenue. & One suggestive quasi-experiment and a few industry claims; causality not established. \\
Rejected as a general claim & ``GEO increases visibility by 40\%.'' & The figure is a relative maximum on one metric under a specific configuration. \\
\bottomrule
\end{tabularx}
\end{table*}

The most important conclusion concerns scope. The evidence is strong for a causal effect conditional on context, moderate for certain informational properties, and weak for transmission through to traffic. This gradient explains why practitioners may observe successful cases even though the literature does not warrant a general promise.

The second conclusion concerns objectives. Responsible optimization does not maximize only $C_s$ or $P_s$. It seeks a Pareto frontier among discoverability, utility, fidelity, cost, stability, and fairness. FeatGEO and several agentic approaches move in this direction, but their metrics remain largely automated \citep{liu2026featgeo,yuan2026agenticgeo}. The future unit of evaluation should be a multi-objective policy subject to truthfulness constraints, not a text that wins a citation contest.

\section{Recommended Protocol for Reproducible GEO Measurement}

\subsection{Define the Estimand Before Data Collection}

A study must state whether it targets: (i) the conditional effect of a rewrite once the document has been injected; (ii) the total effect in a reproducible pipeline; (iii) an observational association with a commercial surface; or (iv) a business outcome in production. Conflating these estimands yields invalid conclusions. The primary metric, denominator, exclusions, and threshold for a substantively meaningful difference must be prespecified.

\subsection{Minimum Factorial Design}

We recommend crossing the following factors:
\begin{itemize}
  \item multiple named engines and search modes, with version and date;
  \item multiple intents and domains, analyzed separately;
  \item three to five paraphrases per information need;
  \item closely spaced repetitions and multiple time windows;
  \item an untreated baseline, an intervention, and, where possible, a placebo of comparable length;
  \item randomized or counterbalanced context order;
  \item multiple adoption rates when competitors are present.
\end{itemize}

Seven to eight repetitions constitute a reasonable starting point based on \citet{schulte2026measure}, but the pilot study should estimate the variance. The main data collection can then be powered for the desired confidence interval. Outputs without search, without citations, or with errors are outcomes, not data to be discarded.

\subsection{Statistical Analysis}

Queries, sources, and dates are clustered units. Testing all generations as though they were independent underestimates uncertainty. For a binary citation outcome, one possible hierarchical model is
\begin{equation}
\operatorname{logit}\Pr(C_{iqetr}=1)=\beta_0+\beta_1 T_i+b_q+b_e+b_t+b_s+\gamma^{\top}X,
\end{equation}
where $b_q,b_e,b_t,b_s$ capture query, engine, date, and source, and $X$ contains the prespecified factors. For proportions or ranks, suitable models or a cluster bootstrap at the query--source level are preferable. Studies should report absolute effects, relative effects, intervals, and distributions, not merely a mean.

In production, randomization across pages or periods is ideal. When this is infeasible, difference-in-differences, interrupted time series, and synthetic controls can help, but parallel trends and placebo tests must be examined. The study by \citet{watanabe2026traffic} illustrates the importance of a control: a large raw increase may arise from platform growth.

\subsection{Quality Assurance and Reproducibility}

Subject to the applicable terms of use, the protocol should retain prompts, raw responses, citations, search status, timestamps, locale, account type, user agent, and canonicalization rules. A stratified human sample should verify citation attribution, claim support, sentiment, factual preservation under rewriting, and judge errors. Sensitive or protected data can be released as hashes, metadata, and reconstruction scripts.

\begin{table*}[t]
\centering
\small
\caption{Minimum checklist for a GEO study.}
\label{tab:checklist}
\begin{tabularx}{\textwidth}{L{3.1cm}Y L{3.7cm}}
\toprule
Component & Requirement & Failure avoided \\
\midrule
System & Product, mode, model, date, locale, account, and search enabled & Comparing different surfaces under the same name \\
Sample & Intents, languages, paraphrases, and inclusion rule & Generalizing from a handful of prompts \\
Treatment & Before/after text, length, factuality, and randomization & Confounding the effect with added information or order \\
Repetitions & Closely spaced runs and multiple time windows & Treating a one-off score as a stable rank \\
Pipeline & Retrieval, reranking, context, and generation separated & A downstream gain masking an upstream loss \\
Metrics & Vector $\mathbf V_s$, denominators, and null outcomes retained & Conflating citation with discoverability \\
Statistics & Clustering, intervals, and absolute and relative effects & Pseudoreplication and unstable gains \\
Validation & Human sample, blinded judge, agreement, and errors & Circularity in LLM-based judging \\
Competition & Adoption rates and spillover effects & Non-transportable single-actor prediction \\
Downstream & Control or randomization for clicks and traffic & Attributing platform growth to GEO \\
Integrity & Verified sources, disclosure, and no hidden instructions & Optimization that rewards misinformation \\
\bottomrule
\end{tabularx}
\end{table*}

\section{Governance and the Political Economy of Generative Visibility}

GEO redistributes a resource: attention. Three risks follow. First, engines may concentrate citations on a small number of domains and create barriers to entry. Second, sponsored or optimized content may influence an answer without disclosure equivalent to that required in conventional advertising. Third, if generated summaries substitute for clicks, creators may reduce investment in producing original sources.

\citet{wu2026ecosystem} model the trade-off among user experience, citations, content-production effort, and compensation. Their conclusions depend on a theoretical model, but they frame the question correctly: citation is not merely a fidelity metric; it is part of an incentive mechanism. \citet{wen2026risks} argue that governance should address concentration, disclosure, and blind spots between academic research and industry practice.

Several operational principles follow:
\begin{itemize}
  \item disclose commercial relationships that may affect a recommendation;
  \item enable publishers to inspect or challenge incorrect attribution;
  \item publish aggregate audits of source concentration and diversity;
  \item treat indirect instructions as untrusted content by default;
  \item do not reward a citation metric without a factual-support constraint;
  \item study compensation, licensing, and traffic jointly rather than separately.
\end{itemize}

Regulation should target mechanisms and incentives rather than freeze a list of ``GEO-compliant'' formats. Engines evolve too rapidly for a purely technical taxonomy, whereas criteria concerning truthfulness, disclosure, and avenues for redress are more stable.

\section{Research Agenda}

\subsection{Connecting Crawling, Indexing, and Generation}

The priority is a benchmark in which a modified page must actually be crawled, indexed, retrieved, reranked, and then cited. SAGEO Arena provides a reproducible step in this direction, but extending the analysis to commercial systems requires field protocols, indexing delays, and controlled pages. Content modifications, internal linking, structured data, domain reputation, and crawler accessibility must be separated.

\subsection{Measuring Absorption Causally}

Composite absorption scores must be validated against interventions with and without the source, atomic factual units, and, where available, attribution traces provided by the engine. The question is not merely ``which text resembles the page?'' but ``which claim or decision would have changed in its absence?''

\subsection{Observing Human Attention}

The \PAWC assumes that the beginning of an answer attracts more attention. Experiments using eye tracking, clicks, recall, and choice could estimate empirical position weights. They should include mobile responses, spoken output, conversational mode, and the effects of trust induced by a citation.

\subsection{Studying Personalization, Languages, and Geographies}

Most studies use English, anonymous accounts, and a small number of locations. The differences in overlap between the United States and Germany observed by \citet{kirsten2026characterizing} motivate multilingual and multi-region panels. Studies must also distinguish the absence of local sources from engine bias.

\subsection{Treating Drift as an Object of Study}

Rather than treating an engine update as a nuisance, research should measure drift regimes: abrupt changes, seasonality, citation aging, and policy transfer. A living benchmark should version prompts, snapshots, and page content rather than publish a supposedly timeless score.

\subsection{Multi-Actor Experiments and Equilibria}

Studies should randomize adoption rates and examine effects on small publishers, large brands, and public-interest sources. Relevant metrics include individual gain, informational welfare, concentration, optimization cost, and post-equilibrium quality. A strategy that is useful at 10\% adoption but harmful at 100\% cannot be recommended without qualification.

\subsection{Defenses and Integrity Certification}

Research on attacks must be coupled with responsible-disclosure constraints and defenses evaluated for false positives. A standard could certify that a transformation preserves facts, retains sources, adds no model-directed instructions, and remains readable to users. Certification would not guarantee a rank; it would guarantee the integrity of the optimization process.

\subsection{Connecting Visibility to Value}

Partnerships with publishers and engines are necessary to connect $D_s$, $C_s$, $H_s$, and $B_s$. Randomized trials should measure not only clicks but also traffic quality, conversion, satisfaction, and substitution effects. Without this link, GEO risks optimizing decorative citations.

\section{Reproducibility Artifacts}

The accompanying arXiv source package contains the LaTeX manuscript, the BibTeX bibliography, a detailed CSV matrix of the 45 studies, and a retrospective CSV record of the databases, query families, deduplication rule, and review limitations. The original database-specific hit counts and complete exclusion ledger were not retained and are explicitly marked as unavailable. A dated copy should also be archived in a persistent repository because the publication status of the preprints will change.

\section{Conclusion}

Between November 2023 and July 2026, GEO evolved from a set of heuristics into a research program on source selection, attribution, absorption, and competition. The foundational paper correctly identified a new visibility surface and demonstrated that the presentation of an already retrieved document can alter its share of an answer. Subsequent work has, however, narrowed the scope of that conclusion: general heuristics transfer poorly, context order and relevance often dominate, competition can erode gains in tested multi-actor settings, and downstream optimization can impair retrieval.

Commercial audits likewise show that engines cite different source ecosystems, vary across runs, and may attribute claims to pages that do not adequately support them. Visibility therefore cannot be reduced to ``being cited by ChatGPT.'' It must be measured as a vector across multiple engines, prompts, and periods, while retaining fidelity measures and null outcomes.

Within the reviewed corpus, the best-supported synthesis is both robust and limited: \emph{already retrieved content can causally influence an answer, while the review identified no technique with a stable, longitudinal, cross-platform causal effect on organic discoverability or downstream clicks and conversions}. This limitation does not invalidate GEO. It defines the scientific work still required to turn it into a cumulative discipline rather than a promotional vocabulary.

\clearpage
\onecolumn
\appendix

\section{Evidence Hierarchy}

\begin{table}[htbp]
\centering
\small
\caption{Proposed framework for grading a GEO claim.}
\begin{tabularx}{\textwidth}{L{1.0cm}L{3.2cm}YY}
\toprule
Level & Design & Example of a warranted conclusion & Caveat \\
\midrule
A & Randomized field trial or strong quasi-experiment with logs and controls & Causal effect on clicks, traffic, or conversions in the setting studied & Limited transportability across products and periods \\
B & Live commercial engines, repetitions, paraphrases, and multiple dates & Distribution of external visibility and cross-engine differences & API/interface configuration and personalization may be incompletely characterized \\
C & Commercial system with manually supplied URLs or files & Black-box post-retrieval effect & Does not establish crawling, indexing, or organic retrieval \\
D & Reproducible RAG pipeline with retrieval, reranking, and generation & Stage-specific causal effects and the total effect within the testbed & External validity for proprietary products \\
E & Fixed context, synthetic ranker, or LLM judge & Possible mechanism conditional on the context & Weak evidence of real-world web visibility \\
\bottomrule
\end{tabularx}
\end{table}

A study may combine several levels. For example, \citet{pfrommer2024ranking} include both controlled experiments and post-retrieval commercial validation, while \citet{watanabe2026traffic} observe real logs but obtain causal identification that is limited by the placebo test. The level grades a specific claim, not the overall value of a paper.

\section{Condensed Matrix of Core Studies}

\footnotesize
\begin{longtable}{L{2.55cm}L{1.8cm}L{3.05cm}L{8.0cm}}
\caption{Directly relevant studies included in the review. \peer{} = peer-reviewed; \forth{} = accepted and forthcoming as of the cutoff date; \preprint{} = preprint; \workshop{} = workshop.}\label{tab:corpus}\\
\toprule
Study & Status & Main stage & Key result or limitation \\
\midrule
\endfirsthead
\multicolumn{4}{l}{\textit{Table~\thetable{} (continued)}}\\
\toprule
Study & Status & Main stage & Key result or limitation \\
\midrule
\endhead
\midrule
\multicolumn{4}{r}{\textit{Continued on next page}}\\
\endfoot
\bottomrule
\endlastfoot
Aggarwal et al. 2024 & \peer & citation/prominence & Up to approximately 40\% relative improvement in a five-document context; no end-to-end retrieval. \\
Liu et al. 2023 & \peer & fidelity & 51.5\% of sentences fully supported and 74.5\% of citations correct across four historical engines. \\
Li and Sinnamon 2024 & \peer & commercial sources & 1\,008 responses overall; 26\% domain overlap in the 672-response Bing--Perplexity phase. \\
Wan et al. 2024 & \peer & influence & Topical relevance dominates in ConflictingQA; controlled context. \\
Kumar and Lakkaraju 2024 & \preprint & product ranking & Strategic sequences effective in a fictitious catalog; no commercial engine. \\
Pfrommer et al. 2024 & \peer & post-retrieval attack & Injection transfers to Perplexity; URLs explicitly supplied. \\
Nestaas et al. 2025 & \peer & commercial attack & Preference manipulation on Bing, Perplexity, and plugins; controlled pages. \\
Narayanan Venkit et al. 2025 & \peer & use/verifiability & Qualitative study showing the practical limitations of citations; 3 pilot and 21 main-study participants. \\
Puerto et al. 2025 & \peer & competition & Three positive cases out of 54; none in QA; gains approach zero under broad adoption. \\
Qian et al. 2025 & \peer & adversarial ranker & Exploitable decision-making blind spot in LLM ranking. \\
Zheng et al. 2025 & \peer & defense & GRADA substantially reduces attack success with limited accuracy loss. \\
Wu et al. 2026 (AutoGEO) & \peer & learned optimization & Reported average improvement of +35.99\%, conditional on five retrieved documents. \\
Chen et al. 2025 (CC-GSEO) & \preprint & influence/quality & Retrieved web contexts, but LLM-generated article-centric queries and predominantly automated judges; no complementary human validation. \\
Chen et al. 2025 (audit) & \preprint & commercial engines & Earned media overrepresented; observational snapshot. \\
Lüttgenau et al. 2025 & \preprint & rewriting & Gains on synthetic travel content; small extrinsic evaluation. \\
Ho et al. 2025 & \workshop & ad retrieval & PPO improves inclusion, with a small absolute MRR gain; offline pipeline. \\
Bagga et al. 2025 & \preprint & e-commerce & Ten of fifteen heuristics neutral or negative; systematically optimized prompts perform better and converge on a domain-agnostic structure. \\
Kim et al. 2026 & \preprint & full pipeline & Downstream rewrites degrade retrieval and reranking; non-commercial pipeline. \\
Tian et al. 2026 & \preprint & citation diagnostics & Targeted repairs reported with 5\% of text modified; corpus inconsistencies. \\
Yuan et al. 2026 & \preprint & multi-objective agents & Offline tests on GEO-Bench, MS MARCO, and a custom Amazon-derived dataset using two open-weight engines and fixed five-document lists. \\
Liu and Xu 2026 & \peer & multi-objective features & Informational features more useful than lexical ones; high cost and substantial reliance on judges. \\
Chen et al. 2026 (Mind Reader) & \peer & latent demand & Large gains, but generation and evaluation are largely performed by LLMs. \\
Zhou et al. 2026 & \peer & multiple queries & Formalizes conflicts and downside risk; main engine is simulated. \\
Wu et al. 2026 (MAGEO) & \peer & agents/attribution & Reusable strategies and a fidelity--visibility score; frozen context. \\
Tang et al. 2025 & \workshop & stealth attack & Rank--fluency trade-off on open models; no commercial engine. \\
Xing et al. 2026 & \peer & ranker attack & Short, naturalistic suffixes promote targets; simplified context. \\
Jin et al. 2026 & \preprint & product ranking & High reported Top-$k$ success; fixed retrieved lists of ten products supplied in JSON and possible fabricated reviews. \\
Nimase et al. 2026 & \preprint & attack benchmark & Compares attacks and white-hat methods on one ranker; shares its name with the original GEO-bench. \\
Smirnov 2026 & \preprint & snippets & RL exploits comparative preferences; simulated overview. \\
Kirsten et al. 2026 & \peer & multi-surface audit & 4\,706 queries, low overlap, and 9--28\% repeated-decision changes on temperature-controllable surfaces. \\
Grossman et al. 2026 & \forth & Google audit & 11\,500 queries; AIO on 51.5\%; cross-surface Jaccard below 0.2. \\
Xu et al. 2026 & \preprint & longitudinal AIO & 55\,393 queries; 13.7\% activation; 11\% of claims insufficiently supported. \\
Allaham and Diakopoulos 2026 & \preprint & synthetic sources & Approximately 16\% of 19\,154 retrieved, classified textual pages; single detector and 27.1\% inaccessible, removed, or non-textual URLs. \\
Schulte et al. 2026 & \preprint & stability & Visibility as a distribution; small Swiss universe. \\
Huang et al. 2026 & \preprint & coverage & Measures the fraction of cited-source content reflected in summary atomic content units; dated Natural Questions dataset. \\
Zhang, He, and Yao 2026 & \preprint & absorption & 21\,143 citations and 72 features; noncausal composite score. \\
Vishwakarma et al. 2026 & \forth & citation factors & 252\,000 trials; relevance and position dominate; two documents injected. \\
Vykopal et al. 2026 & \peer & credibility & Differences in groundedness and questionable sources; restricted domains. \\
Watanabe and Nakayashiki 2026 & \preprint & traffic & Estimated ITS multiplier of 1.82, but placebo $p=0.16$; a single site. \\
Sharma 2026 & \preprint & discovery & Large recognition--discovery gap for 112 startups; two models. \\
Zhang et al. 2026 (Pinterest) & \preprint & production & Reported 20\% production traffic lift versus control; causal attribution insufficiently detailed. \\
Yu et al. 2026 & \preprint & structure & Structural gains reported across six engines; limited validation. \\
Hu 2025 & \workshop & theory/competition & Infinitely repeated prisoner's dilemma for attack choices; conclusions depend on the model. \\
Wu et al. 2026 (ecosystem) & \preprint & economics & Creator competition model in which citations and compensation may sustain effort; model-dependent. \\
Wen et al. 2026 & \peer & governance & Position paper on concentration, disclosure, and academic blind spots. \\
\end{longtable}
\normalsize

\twocolumn
\nocite{chen2025ccgseo,ho2025rewriterank,lewis2020rag,luttgenau2025beyond,smirnov2026biases,tian2026agentgeo,venkit2025search}
\bibliographystyle{plainnat}
\bibliography{refs}

\end{document}